\documentclass{PoS}

\usepackage{graphicx}
\usepackage{amsmath}
\usepackage{amsfonts}
\def\lsim{\raise0.3ex\hbox{$<$\kern-0.75em\raise-1.1ex\hbox{$\sim$}}}
\def\gsim{\raise0.3ex\hbox{$>$\kern-0.75em\raise-1.1ex\hbox{$\sim$}}}
\newcommand{\beq}{\begin{equation}}
\newcommand{\eeq}{\end{equation}}
\newcommand{\beqa}{\begin{eqnarray}}
\newcommand{\eeqa}{\end{eqnarray}}
\def\q{{\boldsymbol q}}

\def\0{{\boldsymbol 0}}

\def\kk{{\kappa}}
\def\pp{{\hat{p}}}

\title{Open charm physics with Heavy Ions: theoretical overview}

\ShortTitle{Open charm physics with Heavy Ions: theoretical overview}

\author{\speaker{Andrea Beraudo}\\
        INFN - Torino\\
        E-mail: \email{beraudo@to.infn.it}}

\abstract{The peculiar role of heavy-flavour observables in relativistic heavy-ion collisions is discussed. Produced in the early stage, $c$  and $b$ quarks cross the hot medium arising from the collision, interacting strongly with the latter, until they hadronize. Depending on the strength of the interaction heavy quarks may or not approach kinetic equilibrium with the plasma, tending in the first case to follow the collective flow of the expanding fireball. The presence of a hot deconfined medium may also affect heavy-quark hadronization, being possible for them to recombine with the surrounding light thermal partons, so that the final heavy-flavour hadrons inherit part of the flow of the medium. Here we show how it is possible to develop a complete transport setup allowing one to describe heavy-flavour production in high-energy nuclear collisions, displaying some major results one can obtain. Information coming from recent lattice-QCD simulations concerning both the heavy-flavour transport coefficients in the hot QCD plasma and the nature of the charmed degrees around the deconfinement transition is also presented. Finally, the possibility that the formation of a hot deconfined medium even in small systems (high-multiplicity p-Au and d-Au collisions, so far) may affect also heavy-flavour observables is investigated.}

\FullConference{VIII International Workshop On Charm Physics\\
		5-9 September, 2016\\
		Bologna, Italy}

\begin{document}

\section{Introduction}
Quantum Chromo-Dynamics is characterized by a non trivial phase-diagram in which, depending on the temperature and the baryon density, the active degrees of freedom may be coloured quarks and gluons or ``white'' hadrons. Relativistic heavy-ion collisions are the only experimental tool to study, under controlled conditions, such a phase diagram. In particular, at the highest energies, nuclear collisions allows one to explore the transition from a plasma of deconfined quarks and gluons (QGP) to a gas of hadrons and resonances in the region of high temperature and (almost) vanishing baryon density. In such a regime, experienced also by the universe during its thermal evolution, the transition is actually a smooth crossover and is accompanied by the spontaneous breaking of chiral symmetry (responsible for most of the baryonic mass of our universe).

The matter produced in heavy-ion collisions undergoes the following evolution. There is an initial pre-equilibrium stage, during which the energy stored in the strong colour-fields is converted into particles (quarks and gluons). There is evidence that, quite rapidly ($\tau_0\,\lsim$ 1 fm/c), the system reaches local thermodynamical equilibrium, spending some time ($\sim 5\!-\!10$ fm/c) in the QGP phase, during which it expands and cools until getting to a regime in which the active degrees of freedom are again mesons and baryons. When the collision rate becomes too small to maintain kinetic equilibrium hadrons decouple and follow a free-streaming until being detected. 
Hence, information on the possible onset of deconfinement and on the properties of the produced medium  has to be obtained quite indirectly, starting from hadronic observables.

\emph{Soft probes}, i.e. low-$p_T$ hadrons (the bulk of particle production), provide evidence for the collective behaviour of the produced system, which displays a hydrodynamics expansion driven by pressure gradients. The success of hydrodynamics in explaining peculiar features of soft-particle production (in particular the azimuthal asymmetry of the spectra) suggests that the medium formed in heavy-ion collisions is characterized by a mean-free-path much smaller than its size, $\lambda_{\rm mfp}\ll L$, and that during most of its expansion it remains close to local thermal equilibrium.
\emph{Hard probes}, on the contrary, are those particles (high-$p_T$ partons, heavy quarks and quarkonia) produced in hard pQCD processes in the very first instants and which, before reaching the detectors, cross the fireball formed in the collision of the two nuclei, performing tomography of the latter. In particular, the suppression of high-$p_T$ particle production, loosely speaking referred to as jet-quenching, provides evidence that the medium formed in heavy-ion collisions is very opaque: partons propagating through it lose a non-negligible fraction of their energy.
Within this framework, heavy-flavour (HF) particles play a peculiar role. Soft observables are described by hydrodynamics, assuming to deal with a system close to local thermal equilibrium as a working hypothesis. The study of jet quenching involves modeling the energy-degradation of \emph{external} probes.
On the other hand, the description of HF observables requires dealing with a more complex situation, developing a setup (based on \emph{transport theory}) able to describe how particles, initially shot into the medium, asymptotically approach kinetic equilibrium with the latter. Notice that, at high-$p_T$, the interest in HF particles is no longer related to their possible thermalization, but to the mass and colour-charge dependence of parton energy-loss and jet quenching: we will not address this issue here.

Why are $c$ and $b$-quarks considered heavy, based on their mass $M$? First of all because $M\gg\Lambda_{\rm QCD}$, setting a hard scale which ensures that their initial production is well described by pQCD. Secondly, because $M\gg T$, entailing that their thermal abundance in the plasma would be negligible: in heavy-ion collisions, with an expanding fireball with a quite short lifetime $\sim 10$ fm/c, the final multiplicity is set by the initial hard production. Finally, because $M\gg gT$, $gT$ being the typical momentum exchange in the collisions with the plasma particles, implying that many soft scatterings are necessary to change significantly the momentum and trajectory of a heavy quark: this will be of relevance in order to developed a simplified transport setup.

The present contribution is organized as follows. In Sec.~\ref{sec:lQCD} we discuss what lattice-QCD calculations tell us about the behaviour of charm around deconfinement. In Sec.~\ref{sec:HI} we give an overview on medium modification of HF observables in heavy-ion collisions, illustrating the complex setup needed to get to a meaningful theory-to-experiment comparison and discussing the information one can obtain from the present experimental data. In Sec.~\ref{sec:pA} we move to smaller systems, discussing whether also in high-multiplicity p-A or d-A collisions one can expect final-state effects associated to the formation of a hot deconfined medium and whether this can affect also HF observables. Finally, in Sec.~\ref{sec:conlusions}, we draw our conclusions and discuss the perspectives of future studies.

\section{Charm degrees of freedom around deconfinement: lattice-QCD results}\label{sec:lQCD}
\begin{figure}[h]
\centering
\includegraphics[height=5.3cm,clip]{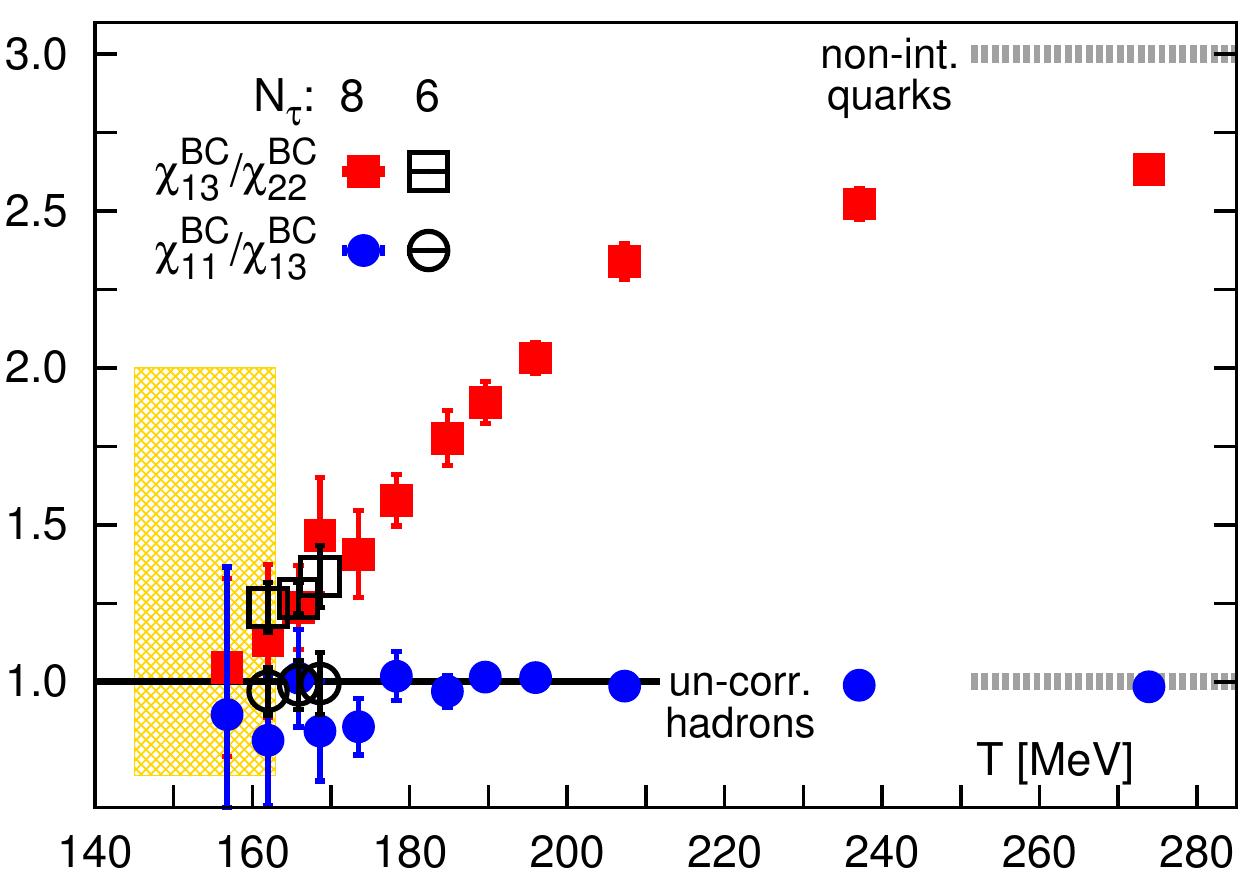}
\includegraphics[height=5.3cm,clip]{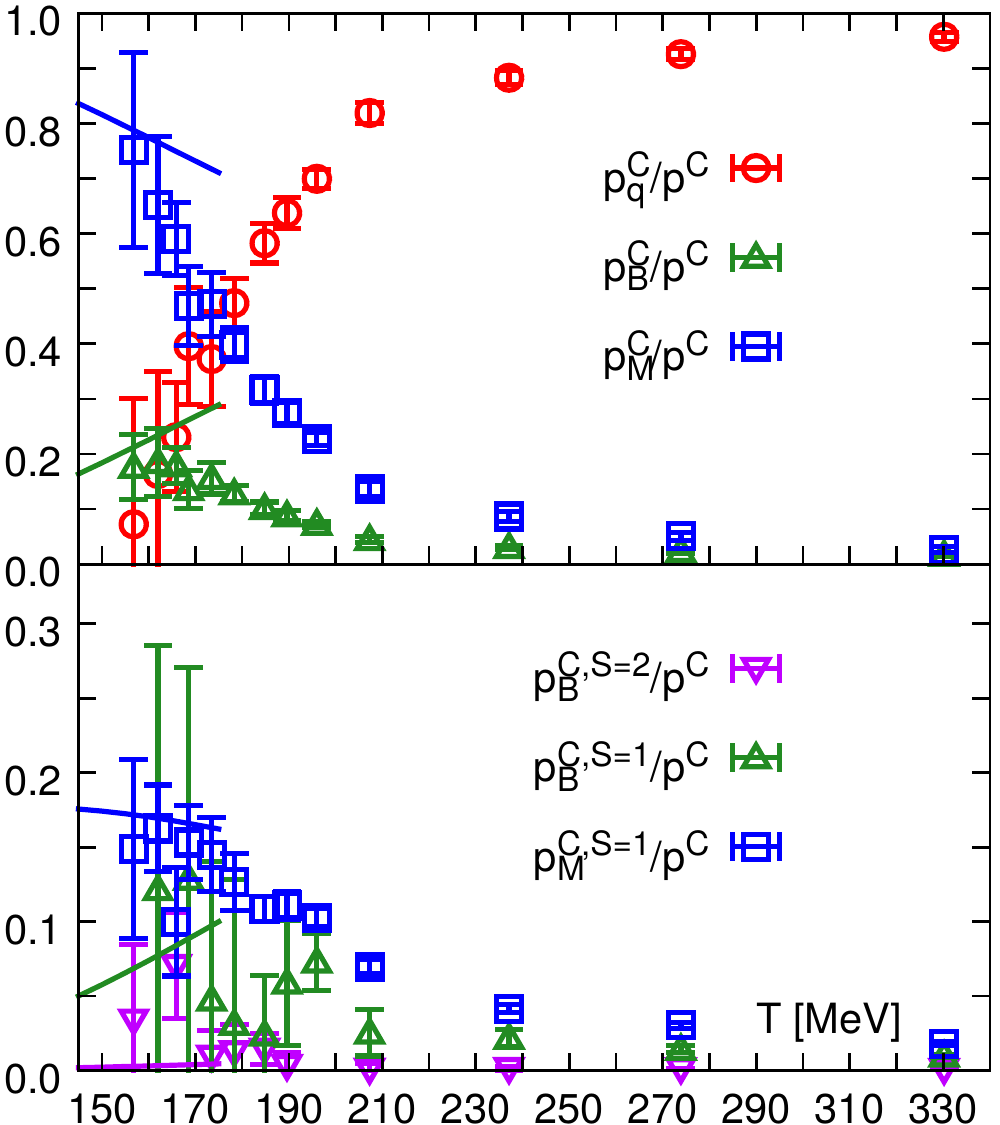}
\caption{Left panel: correlators of charm and baryon-number fluctuations around deconfinement~\cite{lat1}. Right panel: the partial pressures of charmed quarks, mesons and baryons around deconfinement~\cite{lat2}.}
\label{fig:lQCD}
\end{figure}
Lattice-QCD calculations show that, in the limit of vanishing baryon density, the deconfinement transition is actually a crossover, i.e. a rapid but smooth change in the nature of the active degrees of freedom, carriers of conserved charges: something similar to what occurred to the electromagnetic plasma in the primordial universe, with the Saha equation describing the evolution of the ionization fraction, i.e. the fraction of charged protons not bound to electrons in neutral hydrogen atoms. Due to its large mass compared to the other quarks, it is interesting to study the behaviour of charm across the deconfinement transition.
QCD interactions do not change flavour, so that charm can be considered a
conserved charge in heavy-ion collisions, as baryon number. Based on the fact
that charm quarks have $B\!=\!1/3$  while charmed hadrons have $B=0,1$, from
the correlations of $C$ and $B$ fluctuations one can get information on the nature of the charm-carrying degrees of freedom, i.e. whether they are mostly partonic or hadronic ($C$ excess being always associated to $B$ excess in the partonic phase). Taking derivatives of the pressure with respect to the chemical potentials one obtains the various higher-order cumulants of the above fluctuations
\beq
\chi^{BC}_{kl}=\left.\frac{\partial^{k+l}[P/T^4]}{\partial\hat\mu_B^k\partial\hat\mu_C^l}\right|_{\hat\mu_i=0}\quad{\rm where}\quad \hat\mu_i\equiv\mu_i/T,
\eeq
with a very different behaviour in the two phases.
Such an analysis was carried out in~\cite{lat1} and results are displayed in the left panel of Fig.~\ref{fig:lQCD}. It is also possible to separate, as done in~\cite{lat2}, the various partial pressures of charmed particles (quarks and hadrons) around the deconfinement crossover: results are shown in the right panel of Fig.~\ref{fig:lQCD}. Overall, lattice-QCD calculations show that also for charm the deconfinement transition is actually a smooth crossover, with part of the charmed degrees of freedom being still/already hadronic states at temperatures slightly above $T_c$.

\section{Heavy Flavour in heavy-ion collisions}\label{sec:HI}
A realistic study of HF observables in high-energy nuclear collisions requires developing a multi-step setup. First of all, one needs to simulate the initial hard production of the $c\overline{c}$ and $b\overline{b}$ pairs in the elementary nucleon-nucleon collisions: for this automated pQCD tools are available, which one can generalize to include initial-state effects, such as nuclear parton distribution functions (nPDF's) and transverse-momentum broadening in nuclear matter. Secondly, one must have at his disposal a realistic description of the fireball in which the heavy quarks propagate, provided by hydrodynamic calculations validated against soft-hadron data. The core of the setup is represented by the modeling of the heavy-quark propagation in the hot plasma: one has to develop a transport code working in the case of an evolving medium, taking advantage of the previous information. Once the heavy quarks reach a fluid cell below the deconfinement temperature they hadronize and one has to model also this stage: can it be described simply via fragmentation functions like in the vacuum or are other mechanisms, such as recombination with light thermal partons, possible? Clearly, medium modification of hadronization in heavy-ion collisions is an item of interest in itself; however, it is at the same time a source of systematic uncertainty if one wants to extract information on the properties of the deconfined medium.
Finally, since experimental data sometimes does not refer directly to $D$ and $B$-mesons, one must also simulate the decay into the relevant channels (e.g. $D\to X\,\nu_e\, e$, $D\to X\,\nu_\mu\, \mu$, $B\to X\, J/\Psi$...) accessible to the experiments.
\subsection{Heavy-quark production}
\begin{figure}[h]
\centering
\includegraphics[height=4cm,clip]{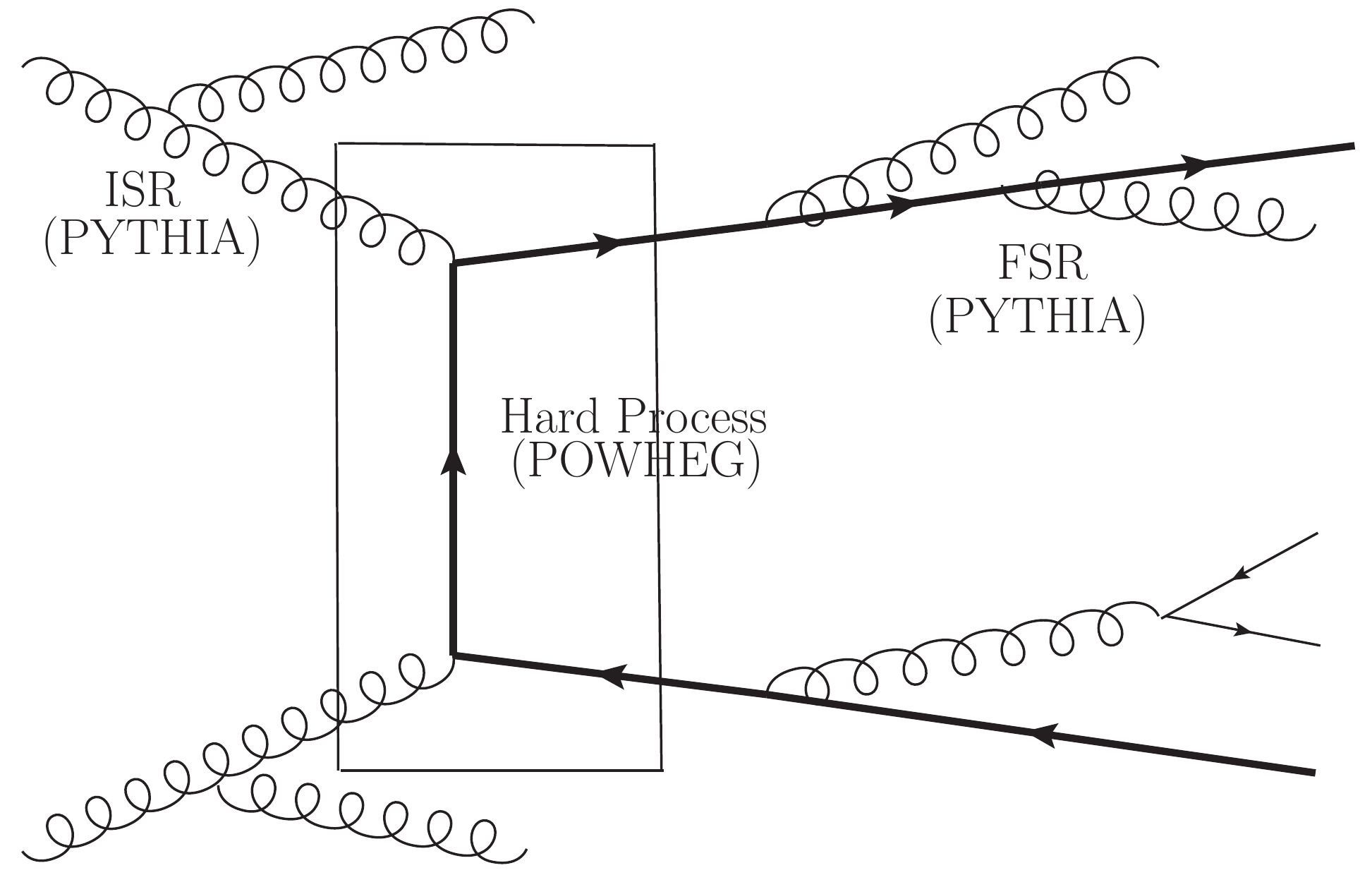}
\includegraphics[height=4.45cm,clip]{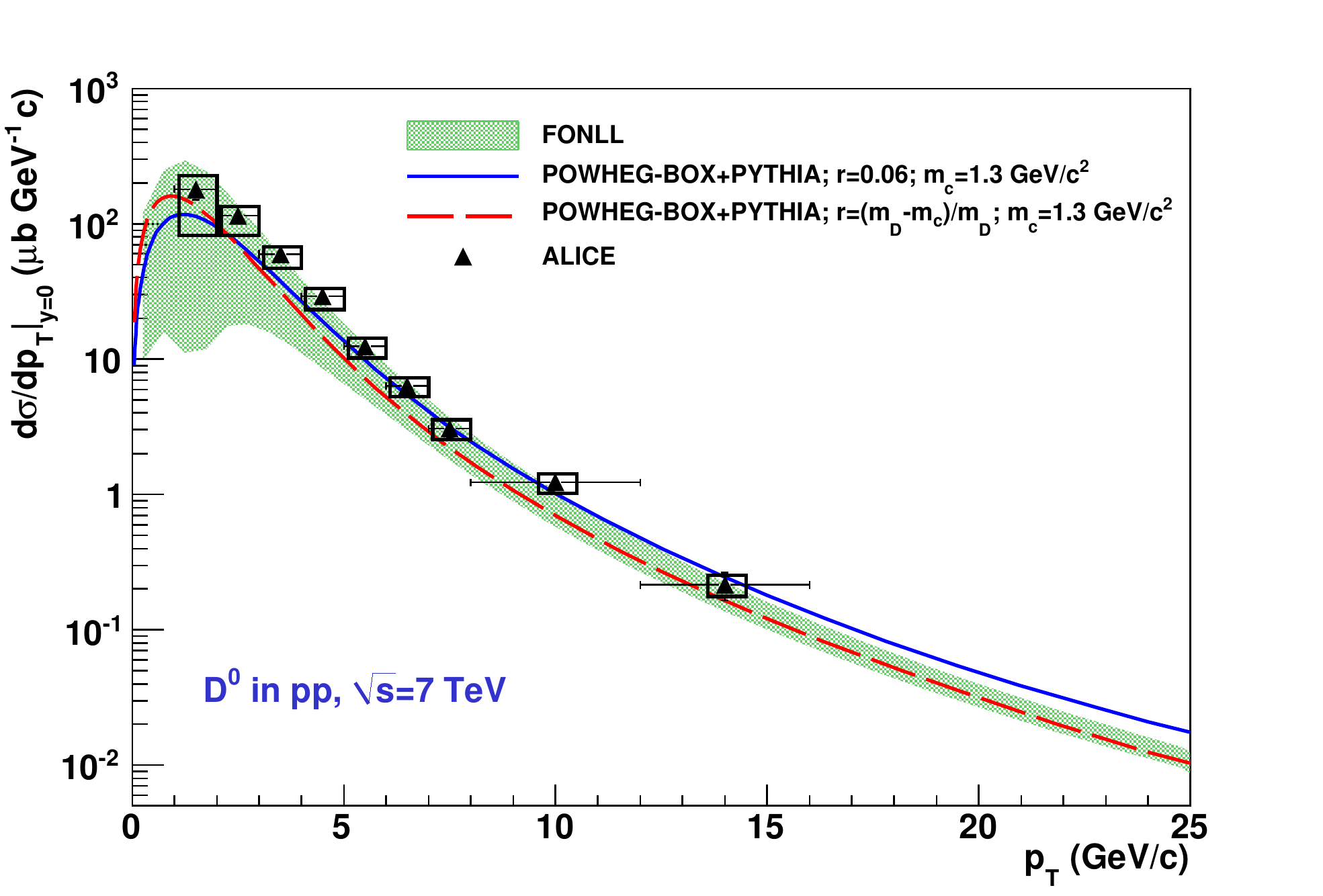}
\caption{Left panel: $Q\overline{Q}$ event generated by POWHEG-BOX~\cite{POWBOX}, with the hard process interfaced with a parton-shower stage. Right panel: FONLL~\cite{FONLL} and POWHEG-BOX predictions (with different fragmentation functions) for $D^0$ production in p-p collisions at 7 TeV  compared to ALICE data~\cite{ALIpp7}.}
\label{hard}       
\end{figure}
Due to the large mass, heavy-quark production is a hard short-distance process occurring at the level of the individual binary nucleon-nucleon collisions. It is then necessary to validate the simulation of the initial hard production against p-p data. The state of the art for its theoretical description is represented by the Fixed Order Next-to-Leading Log (FONLL) calculation, developed to provide accurate results both at low and at high-$p_T$ through a proper resummation of large transverse-momentum logarithms~\cite{FONLL}. However, FONLL simply provides an inclusive single-particle spectrum, while for practical applications it is more convenient to work with an event generator. In this regard, the POWHEG-BOX package~\cite{POWBOX} represents a convenient tool. The latter generates hard events at NLO accuracy and interface them with a parton-shower stage, simulating initial and final-state radiation. 
A comparison of FONLL and POWHEG-BOX results (with $m_c\!=\!1.3$ GeV, $\mu_R\!=\!\mu_F\!=\!m_T$ and HQET fragmentation functions~\cite{Braaten}) with ALICE data for $D^0$ production in p-p collisions at the LHC~\cite{ALIpp7} is displayed in the right panel of Fig.~\ref{hard}. Overall, the data tend to stay at the upper edge of the (quite large, in the case of charm) theoretical uncertainty band, arising from the choice of the quark mass and of the factorization and renormalization scales.
Once validated in the p-p case and supplemented with nPDF's~\cite{eps}, the above pQCD tools can be employed to simulated the initial $Q\overline{Q}$ production in heavy-ion collisions. For each event the full kinematic information on the produced $Q\overline{Q}$ pairs can be stored and used in the subsequent steps of the simulations.
\subsection{Heavy-quark transport}
As a next step, one needs to implement a numerical transport calculation describing the heavy-quark propagation in the plasma.
The starting point of all transport calculation is the Boltzmann equation for the evolution of the heavy-quark phase-space distribution 
\beq
\frac{d}{dt}f_Q(t,\vec x,\vec p)=C[f_Q]\quad{\rm with}\quad C[f_Q]=\int d\q[{w(\vec p+\vec q,\vec q)f_Q(t,\vec x,\vec p+\vec q)}-{w(\vec p,\vec q)f_Q(t,\vec x,\vec p)}],\label{eq:Boltzmann}
\eeq
where the collision integral $C[f_Q]$ is expressed in terms of the $\vec p\to\vec p-\vec q$ transition rate $w(\vec p,\vec q)$. The direct solution of the Boltzmann equation is numerically demanding; however, as long as $q\ll p$ ($q$ being typically of order $gT$), one can expand the collision integrand in powers of the momentum exchange. Truncating the expansion to second order corresponds to the Fokker-Planck (FP) approximation, which leads to the Langevin equation. The latter in its discretized form 
\beq
{\Delta \vec{p}}/{\Delta t}=-{\eta_D(p)\vec{p}}+{\vec\xi(t)},\label{eq:Langevin}
\eeq
provides a recipe to update the heavy quark momentum through the sum of a deterministic friction force and a random noise term specified by its temporal correlator
\beq
\langle\xi^i(\vec p_t)\xi^j(\vec p_{t'})\rangle\!=\!{b^{ij}(\vec p_t)}{\delta_{tt'}}/{\Delta t}\qquad{b^{ij}(\vec p)}\!\equiv\!{\kk_\|(p)}\pp^i\pp^j+{\kk_\perp(p)}(\delta^{ij}\!-\!\pp^i\pp^j).
\eeq
After evaluating the transport coefficients $\kappa_{\|/\perp}(p)$ (representing the average longitudinal/transverse squared momentum exchanged with the plasma per unit time) and $\eta_D(p)$ (fixed by the Einstein relation, so that particles asymptotically reach kinetic equilibrium) one has then to solve Eq.~(\ref{eq:Langevin}) throughout the whole medium evolution~\cite{torino,epj3}, described by hydrodynamic calculations~\cite{rom2,ECHO}.

\begin{figure}[ht!]
\centering
\includegraphics[width=0.42\textwidth,clip]{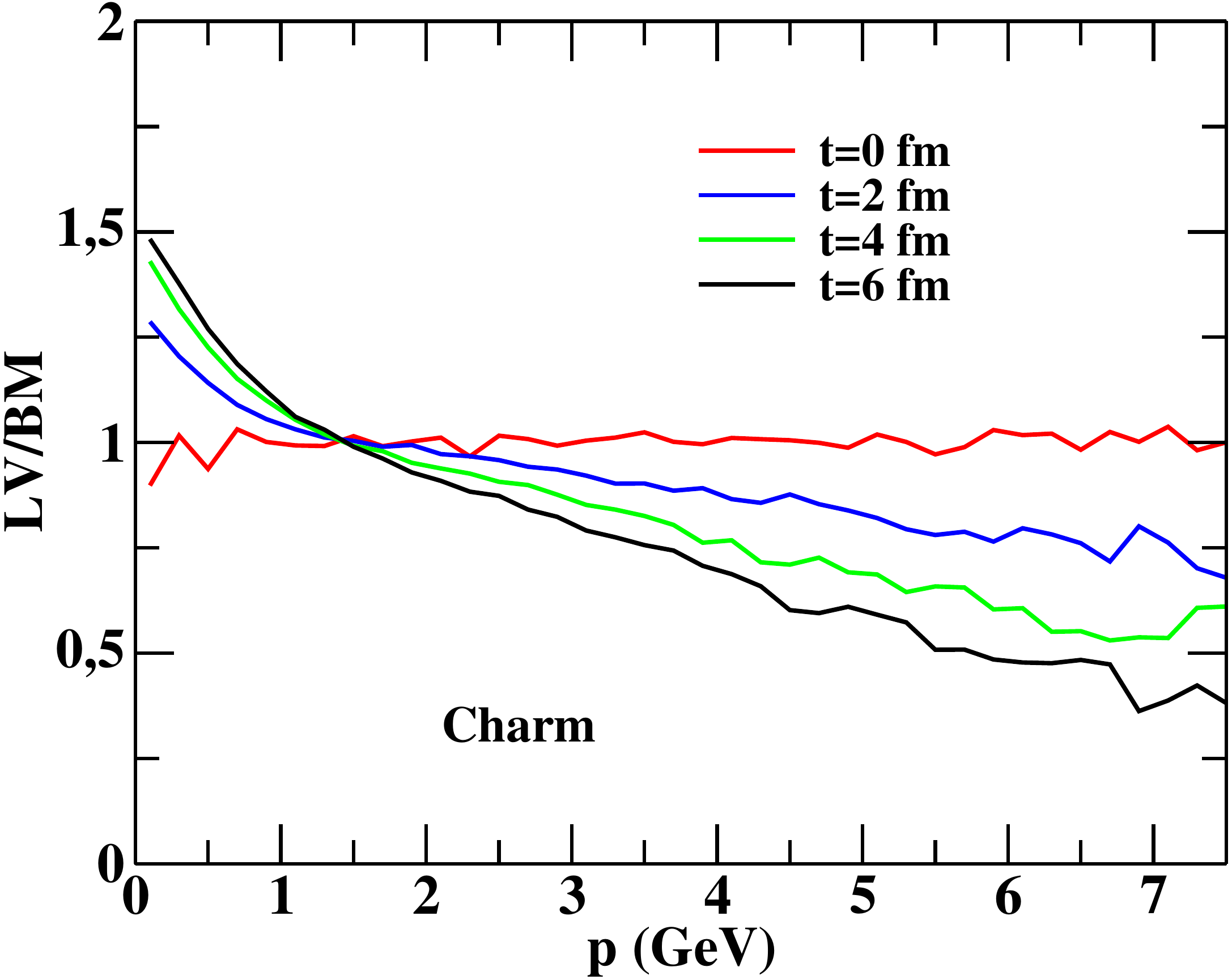}
\includegraphics[width=0.42\textwidth,clip]{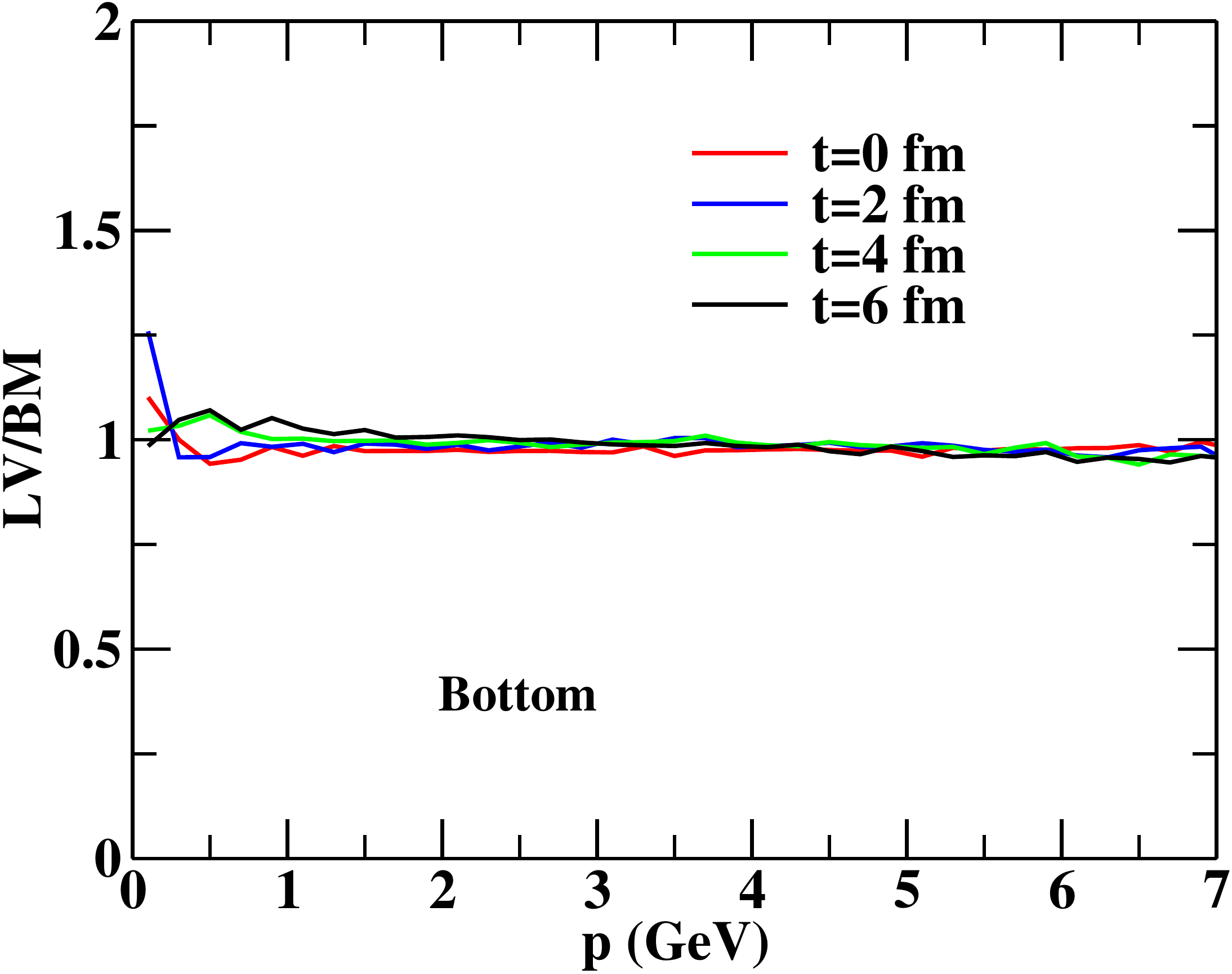}
\caption{Langevin-to-Boltzmann ratio of $c$ (left panel) and $b$-quark (right panel) spectra at different times~\cite{BvsLang}. In the case of beauty the Langevin and the Boltzmann equations provide an equivalent description of the heavy-quark propagation in the medium.}
\label{fig:Boltzmann}
\end{figure}
If the expansion of the collision integral allows one to develop a more tractable transport setup with respect to the original Boltzmann equation, one may wonder whether this soft-scattering approximation is fully justified in situations in which the typical momentum exchange $\sim gT$ is  not so small compared to the heavy-quark mass. The question was addressed in~\cite{BvsLang}, where the authors performed a systematic comparison of the Boltzmann and Langevin approaches: as shown in Fig.~\ref{fig:Boltzmann}, the two schemes provides identical results in the case of beauty, while for charm discrepancies (dependent on the values of medium-parameters) are visible.  

\begin{figure}[h]
\centering
\includegraphics[height=5cm,clip]{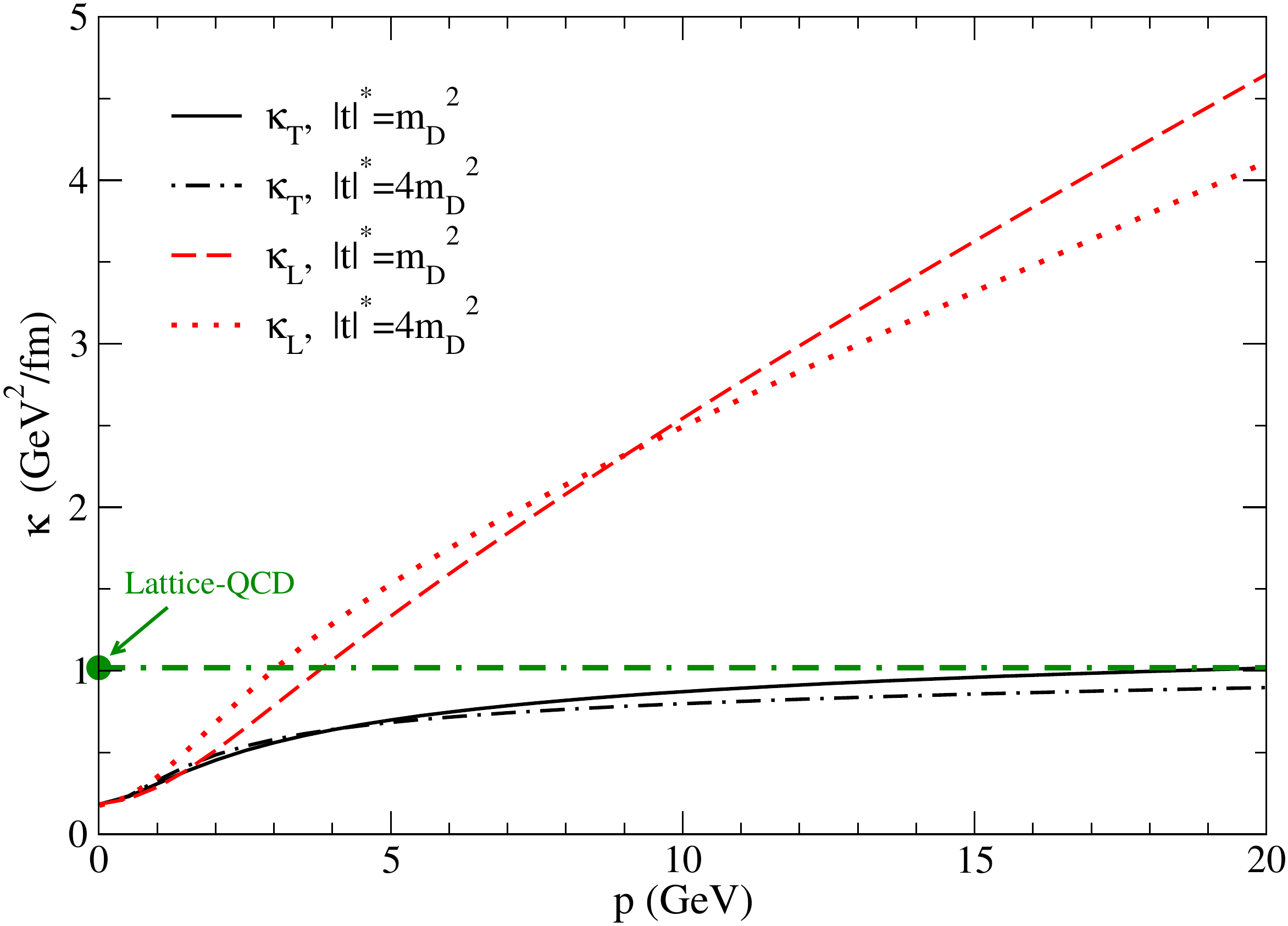}
\includegraphics[height=5cm,clip]{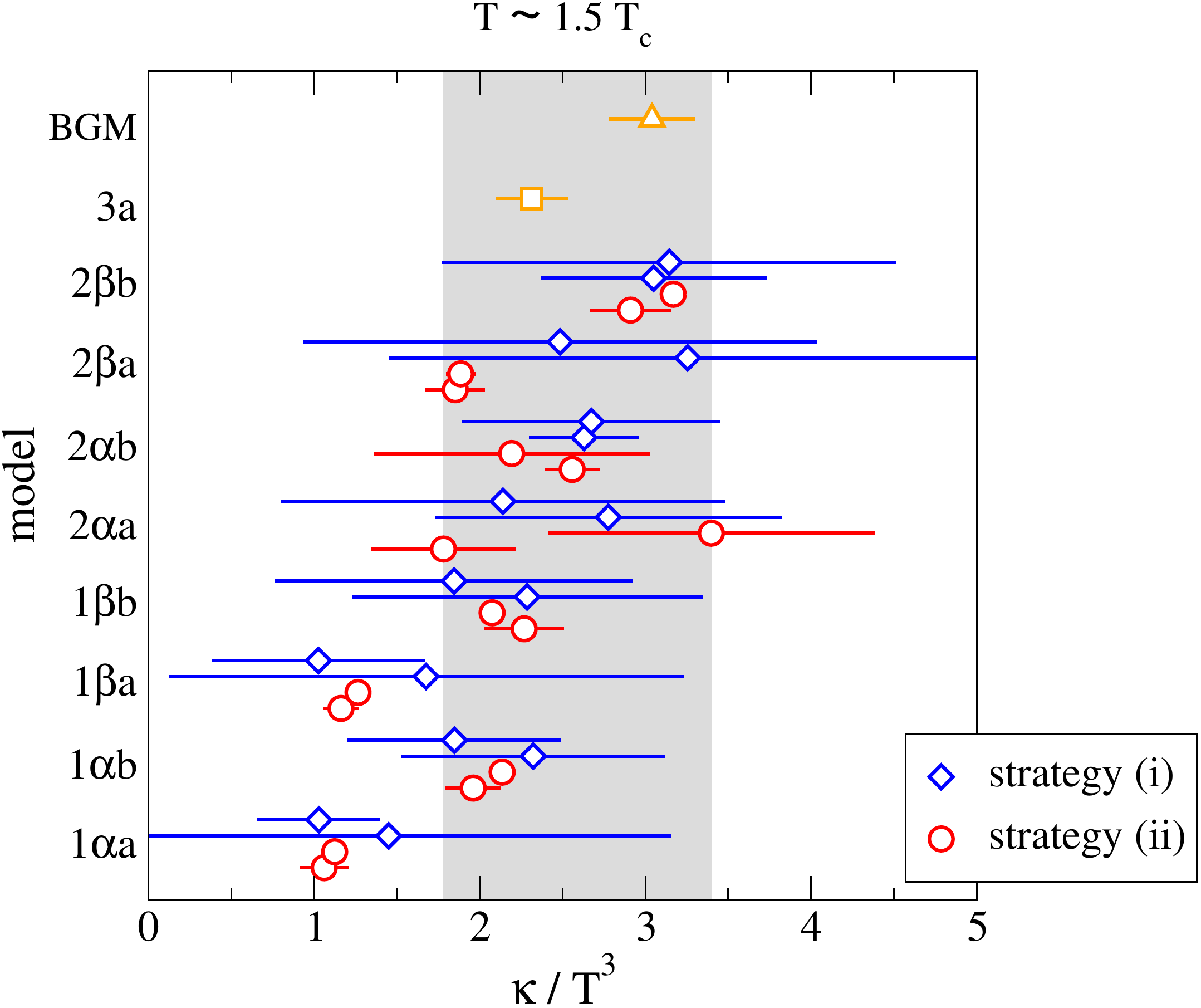}
\caption{Heavy-quark momentum diffusion coefficients. Left panel: weak-coupling results for charm at $T\!=\!400$ MeV, with different choices of the cutoff separating hard and soft collisions~\cite{torino}. Right panel: lattice-QCD result for a static quark ($\kappa_T\!=\!\kappa_L\!\equiv\!\kappa$) in a gluonic plasma~\cite{francis}. The uncertainty band arises from extracting real-time information from imaginary-time correlators.}
\label{kappa}       
\end{figure}
In the Langevin approach physics enters through the transport coefficients, which one aims at deriving through a first-principle calculation. In the left panel of Fig.~\ref{kappa}, referring to charm, we show the results of a weak-coupling evaluation, accounting for $2\to 2$ collisions, with resummation of medium effects in the case of soft-gluon exchange~\cite{torino}. Instead, a non-perturbative estimate can be provided by lattice-QCD simulations, so far limited to the case of a static, infinitely heavy , quark. In this limit one can neglect the momentum dependence of the strength of the noise, so that
\beq
\langle\xi^i(t)\xi^j(t')\rangle\!=\!\delta^{ij}\delta(t-t'){\kappa}\quad\longrightarrow\quad{\kappa}=\frac{1}{3}\int_{-\infty}^{+\infty}\!\!dt\langle\xi^i(t)\xi^i(0)\rangle_{\rm HQ}
\underset{p\to 0}{\sim}\frac{1}{3}\int_{-\infty}^{+\infty}\!\!dt{{\langle F^i(t)F^i(0)\rangle_{\rm HQ}}}.\label{eq:kappa}
\eeq
The momentum-diffusion coefficient $\kappa$ can be extracted from a force-force correlator in an ensemble of states containing one heavy quark: for a static colour source the only force playing a role is the chromo-electric field. Notice that $\kappa$, as any other transport coefficient, is a real-time quantity, to be recontructed from lattice correlators, which, however, can be evaluated only for imaginary-times: this leads to the large theoretical uncertainty band~\cite{francis} in the right panel of Fig.~\ref{kappa}.

Finally, heavy quarks, when reaching a fluid cell below a critical temperature, are allowed to hadronize. Various models have been developed to describe how the medium can modify this stage, all based on the idea that a heavy quark can easily find one or more light partons nearby to recombine with, producing a colour-singlet object, which can be taken to be directly a HF hadron (coalescence~\cite{aic2,rapp}), a heavy resonance~\cite{rapp2} or a string to further fragment~\cite{epj3}.

\subsection{Results}
\begin{figure}[h]
\centering
\includegraphics[width=0.45\textwidth,clip]{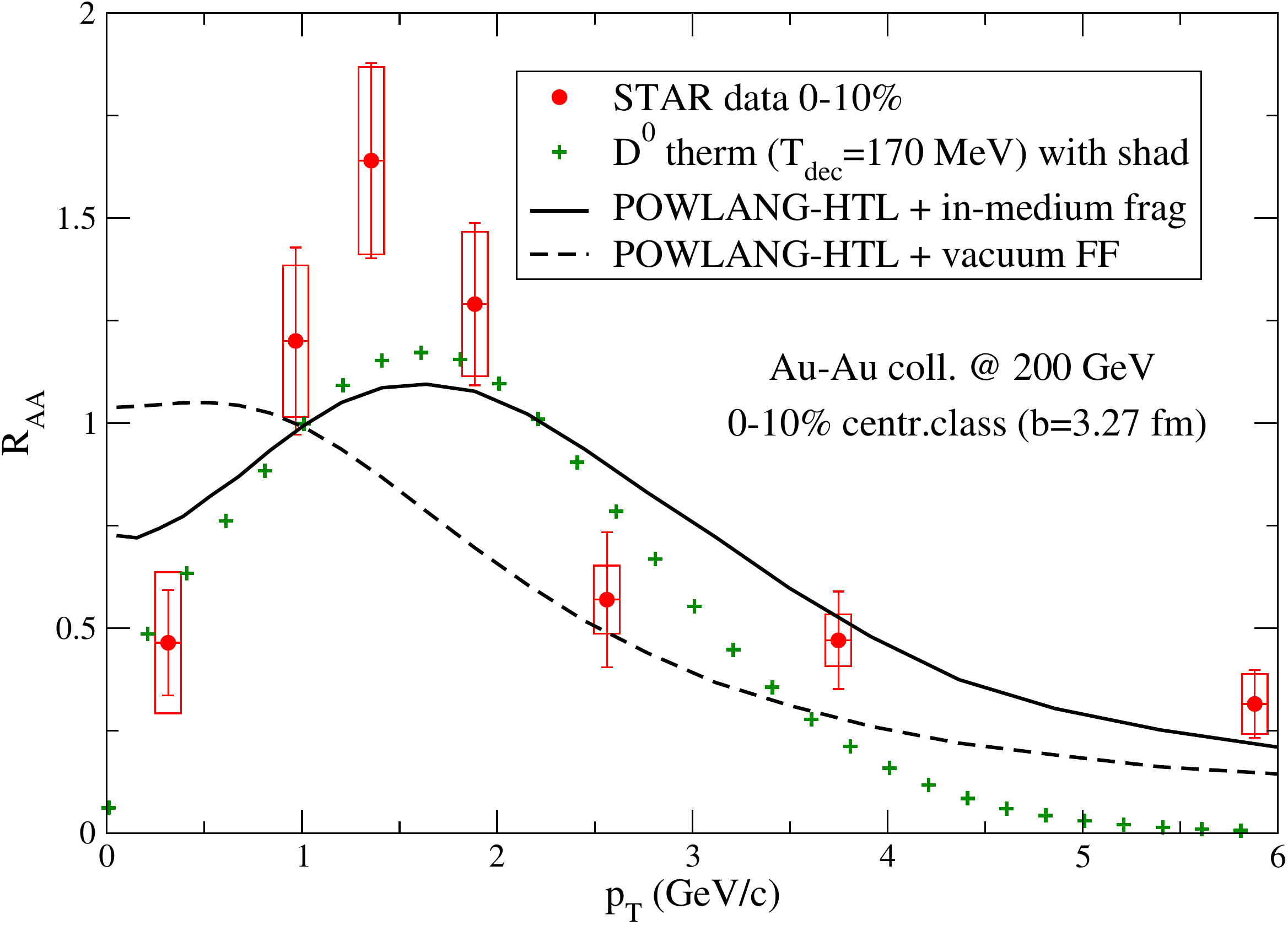}
\includegraphics[width=0.45\textwidth,clip]{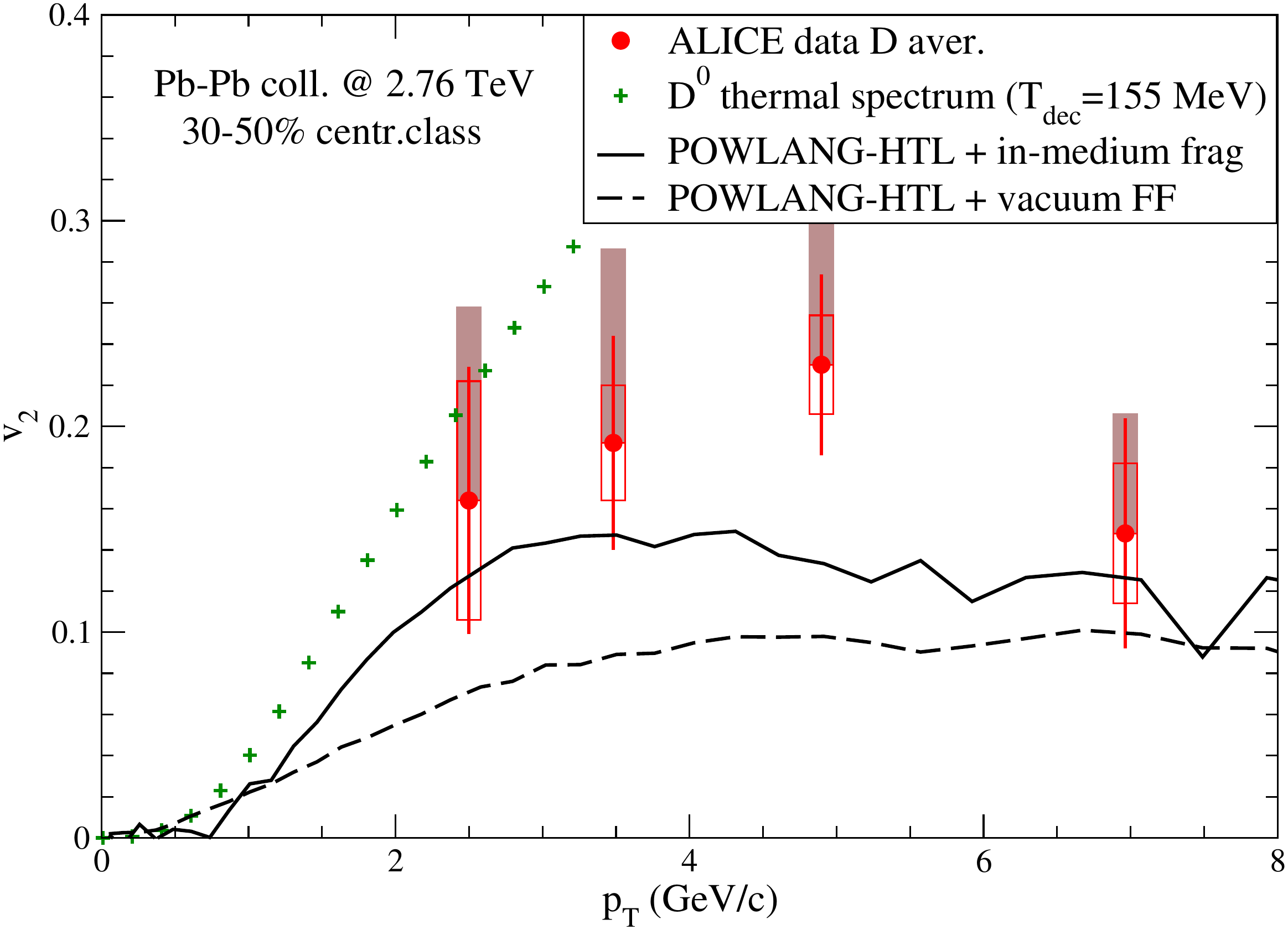}
\caption{Left panel: the nuclear modification factor of $D^0$ spectra in 0-10\% Au-Au collision at $\sqrt{s_{\rm NN}}\!=\!200$ GeV compared to STAR data~\cite{STARD}. Right panel: the elliptic flow of $D^0$ mesons in 30-50\% Pb-Pb collisions at $\sqrt{s_{\rm NN}}\!=\!2.76$ TeV compared to ALICE data~\cite{ALICE_Dv2}. Theory results include the full kinetic-equilibrium limit (green crosses), transport calculations interfaced with vacuum fragmentation functions (dashed curves) and transport calculations followed by in-medium hadronization (continuous curves).}
\label{transpvstherm}       
\end{figure}
Let us now present some selected outcomes of a specific transport setup~\cite{epj3}, pointing out some quite general qualitative findings common with the analysis performed by other groups~\cite{aic2,rapp,rapp2}. In Fig.~\ref{transpvstherm} we consider the \emph{nuclear modification factor} of the $p_T$-spectra and the \emph{elliptic flow}
\beq
R_{\rm AA}\equiv\frac{(dN/dp_T)^{\rm AA}}{\langle N_{\rm coll}\rangle (dN/dp_T)^{\rm pp}}\quad{\rm and}\quad v_2\equiv\langle\cos[2(\phi-\psi_{\rm RP})]\rangle 
\eeq
of $D^0$ mesons in Au-Au and Pb-Pb collisions at RHIC and LHC ($\langle N_{\rm coll}\rangle$ being the average number of binary nucleon-nucleon collisions). It is interesting to start addressing the limit of very (infinitely) large transport coefficients. In this case HF particles would approach local kinetic equilibrium with the fireball and would be thermally emitted from a decoupling hypersurface, identified (rather schematically) by a temperature below which interactions are supposed to become too rare to further modify the spectra. The usual Cooper-Frye formula gives:
\beq
E\frac{dN}{d\vec{p}}=\frac{1}{(2\pi)^3}\int_{\Sigma_{\rm dec}} p^\mu d\Sigma_\mu\,\exp\left[-\frac{p\!\cdot\! u(x)}{T_{\rm dec}}\right].
\eeq
As a consequence, high-energy particles would be thermally suppressed; on the other hand, the collective flow of the fluid $u^\mu$ would tend to drag very soft particles to larger $p_T$. This gives rise to the bump in the $R_{\rm AA}$ seen in the left panel of Fig.~\ref{transpvstherm} (green crosses) and displayed also by the experimental data. At the same time, in non-central collision the flow of the fluid is characterized by an azimuthal anisotropy, due to the initial larger pressure gradient along the reaction-plane direction $\psi_{\rm RP}$: the azimuthal distribution of particles flowing with the fluid would inherit this anisotropy, displaying a strong elliptic flow, as shown in the right panel of Fig.~\ref{transpvstherm} and supported by the data. Does the above considerations entail that in heavy-ion collisions charm actually reaches kinetic equilibrium with the medium during its partonic phase? This is not necessarily the case, as shown by the curves in Fig.~\ref{transpvstherm} referring to Langevin calculations with finite (weak-coupling) transport coefficients. Transport in the QGP phase alone does not appear able to transfer to the charmed particles enough radial and elliptic flow (dashed curves with vacuum fragmentation functions). On the other hand, modeling HF hadronization through recombination with thermal partons from the medium leads to results in better agreement with the data, thanks to the additional flow acquired from the light companion (continuous curves).   

\begin{figure}[h]
\centering
\includegraphics[width=0.48\textwidth,clip]{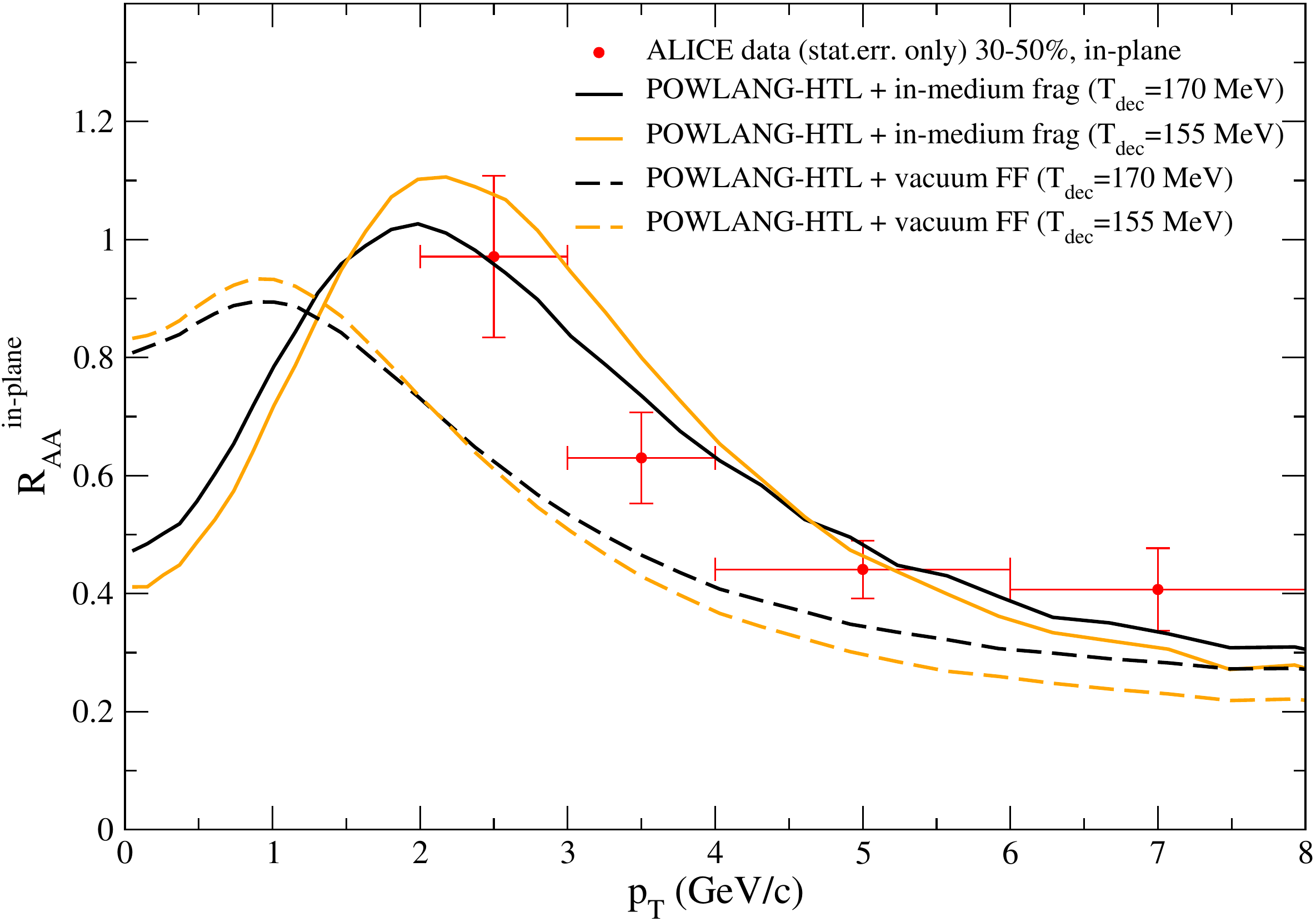}
\includegraphics[width=0.48\textwidth,clip]{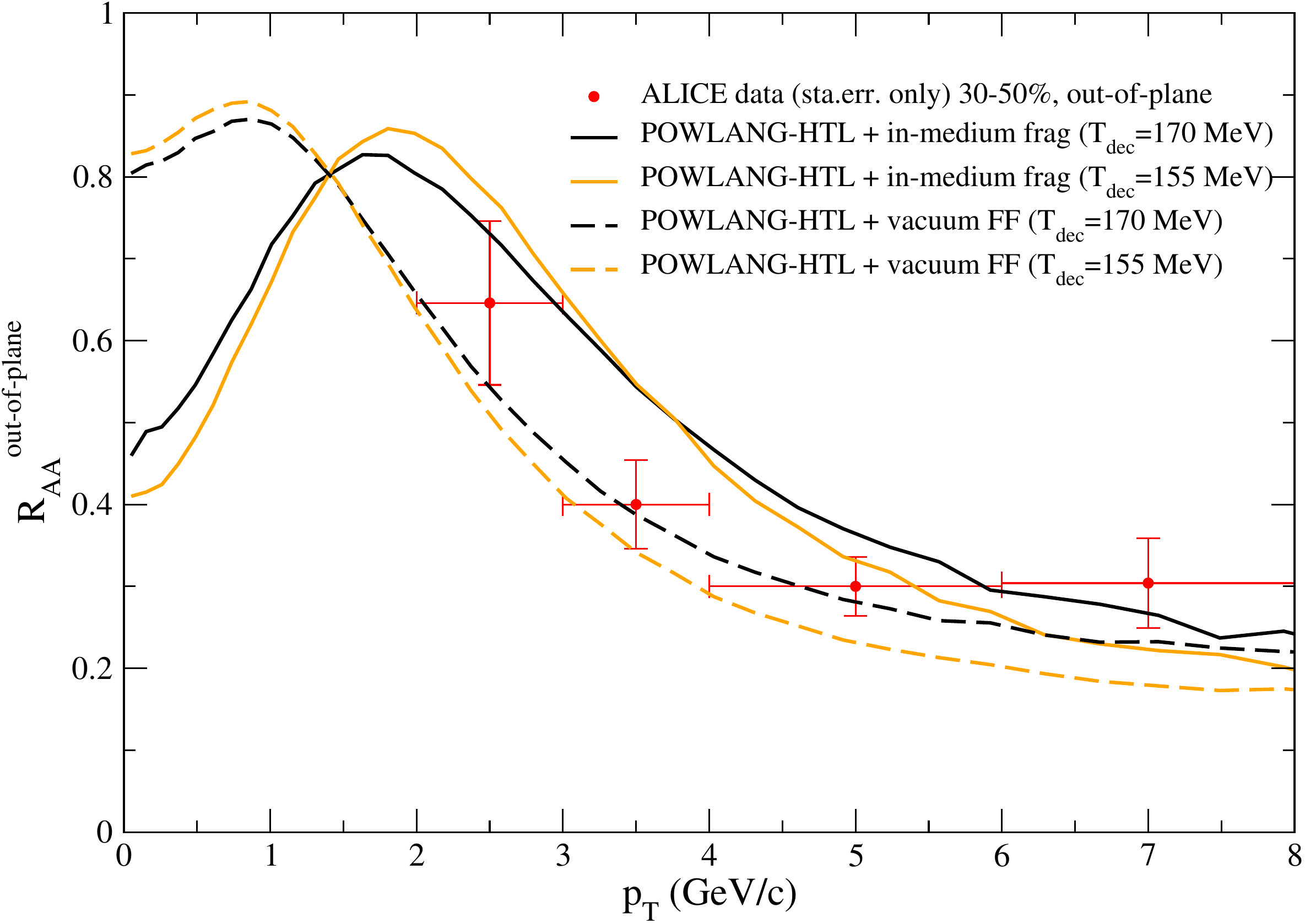}
\caption{The in-plane (left panel) and out-of-plane (right panel) nuclear modification factor of $D^0$ mesons in 30-50\% Pb-Pb collisions at $\sqrt{s_{\rm NN}}\!=\!2.76$ TeV. Transport results with different hadronization schemes and decoupling temperatures are compared to ALICE data~\cite{ALICE_reactionplane}.}
\label{inout} 
\end{figure}
In Fig.~\ref{inout} we compare the nuclear modification factors of $D^0$ mesons in non-central Pb-Pb events at the LHC emitted along (in-plane) or orthogonally (out-of-plane) to the direction of the impact parameter of the collision. We explore the sensitivity to different decoupling temperatures and hadronization schemes. Curves including in-medium hadronization are again characterized by a bump at moderate $p_T$: experimental data at lower $p_T$ are clearly necessary in order to confirm such an occurrence.

\begin{figure}[h]
\centering
\includegraphics[height=4.6cm,clip]{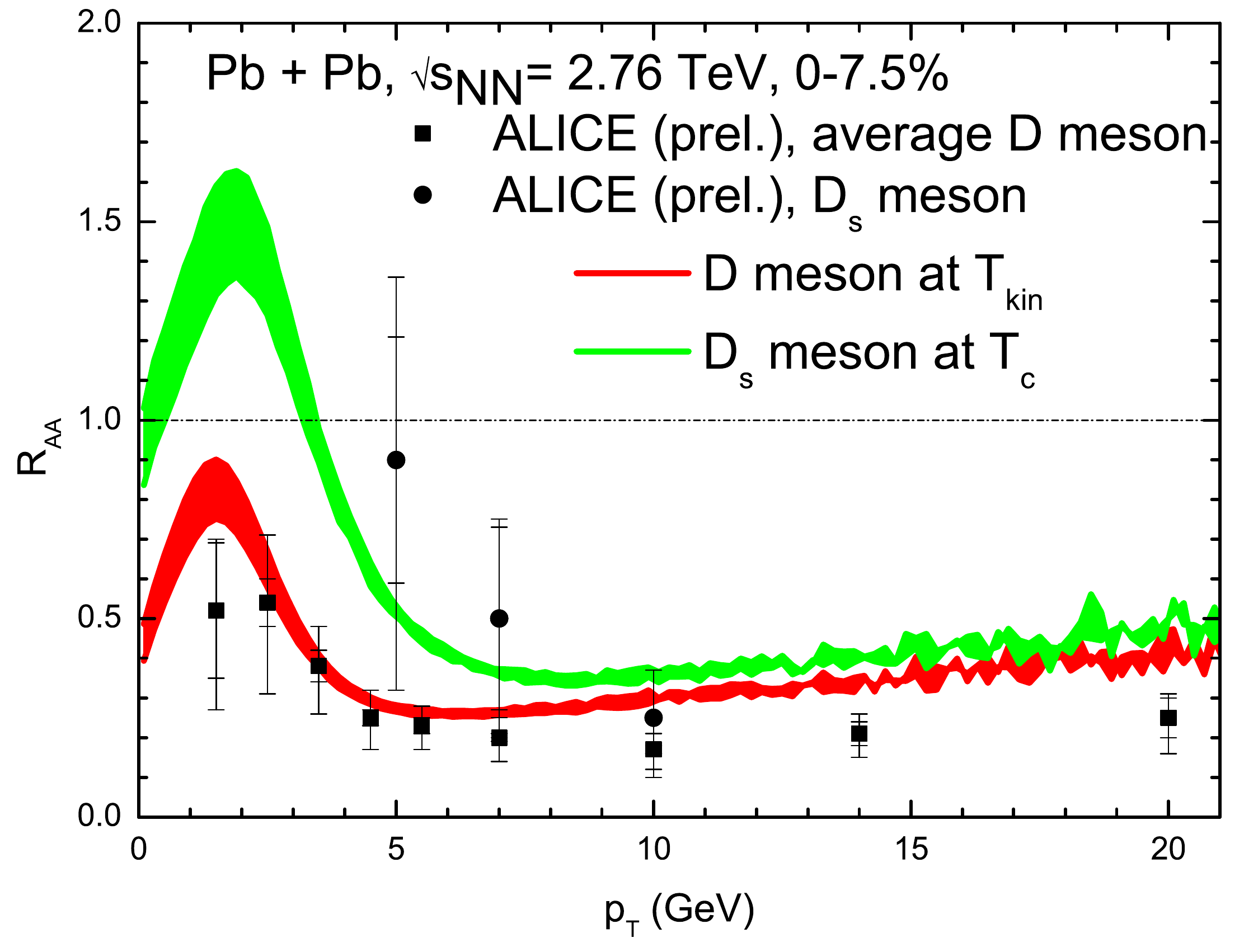}
\includegraphics[height=4.6cm,clip]{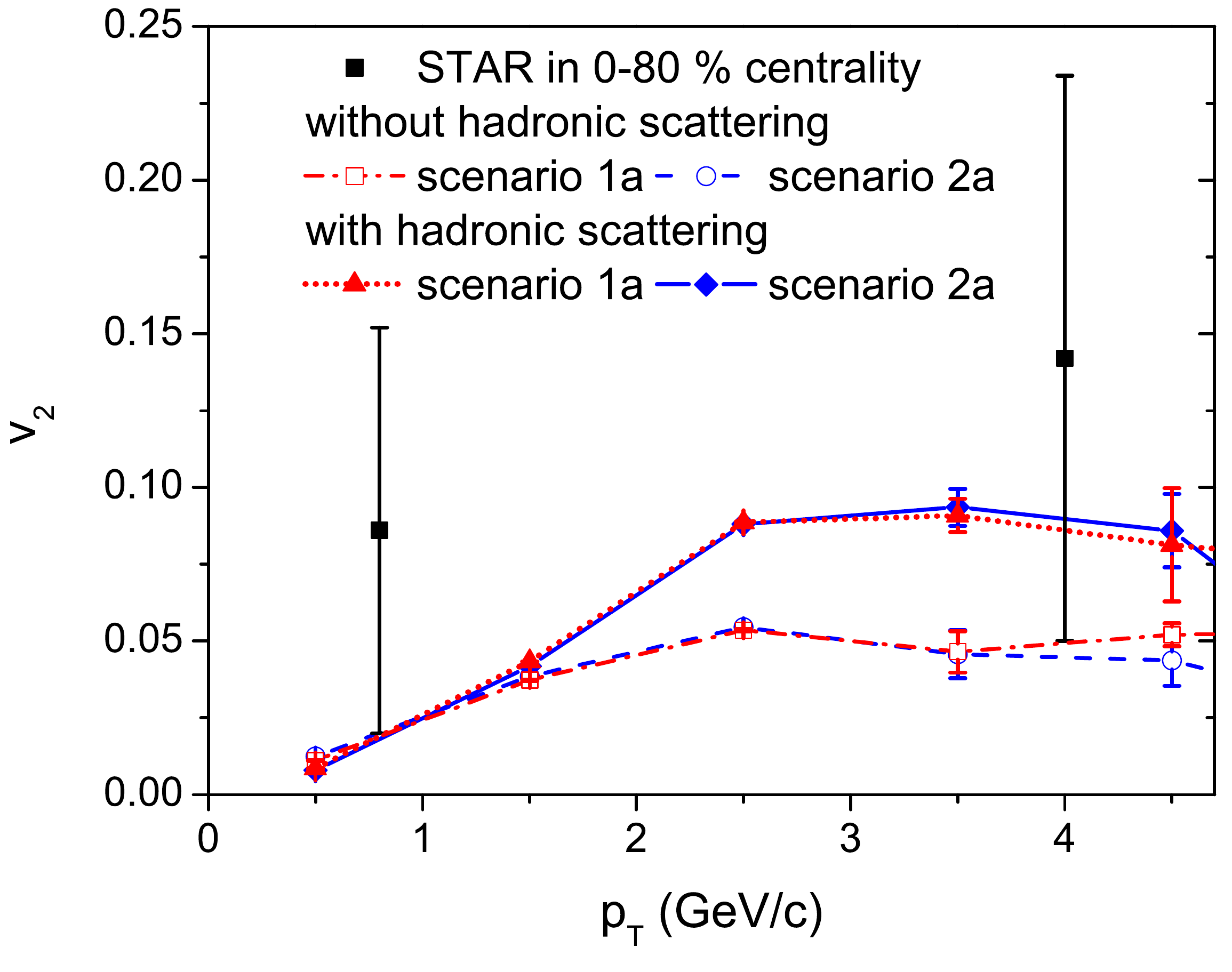}
\caption{Left panel: The nuclear modification factor of $D$ and $D_s$ mesons in central Pb-Pb collisions at $\sqrt{s_{\rm NN}}\!=\!2.76$ TeV; transport results with in-medium hadronization~\cite{he} are compared to ALICE data~\cite{anas}. Right panel: the effect of rescattering in the hadronic phase on the $D$-meson elliptic flow in Au-Au collisions at  $\sqrt{s_{\rm NN}}\!=\!200$ GeV; results of model calculations~\cite{song} are compared to STAR data.}
\label{Ds+HG}
\end{figure}
Besides affecting the momentum and angular distribution of the produced particles, in-medium hadronization may also change the heavy-flavour hadrochemistry. The issue was addressed, for instance, in~\cite{he}, where the authors, implementing a mechanism based on the formation/decay of heavy-light resonances, studied the production of $D$ and $D_s$ mesons in nucleus-nucleus collisions. Results for their nuclear modification factor are displayed in the left panel of Fig.~\ref{Ds+HG}, compared to preliminary ALICE data~\cite{anas}: a relative enhancement of $D_s$ mesons, due to recombination with strange quarks at hadronization, is visible. Finally, also the interactions in the hadronic phase may affect the final particle distributions. In particular, this can be of relevance for the $D$-meson $v_2$. In fact, although transport coefficients are much smaller, in the hadronic phase the medium is characterized by the largest elliptic flow: results obtained in~\cite{song} modeling $D-\pi$ interactions with and effective chiral Lagrangian are displayed in the right panel of Fig.~\ref{Ds+HG}.

\section{Heavy Flavour in small systems: room for medium effects?}\label{sec:pA}
\begin{figure}[h]
\centering
\includegraphics[width=0.45\textwidth,clip]{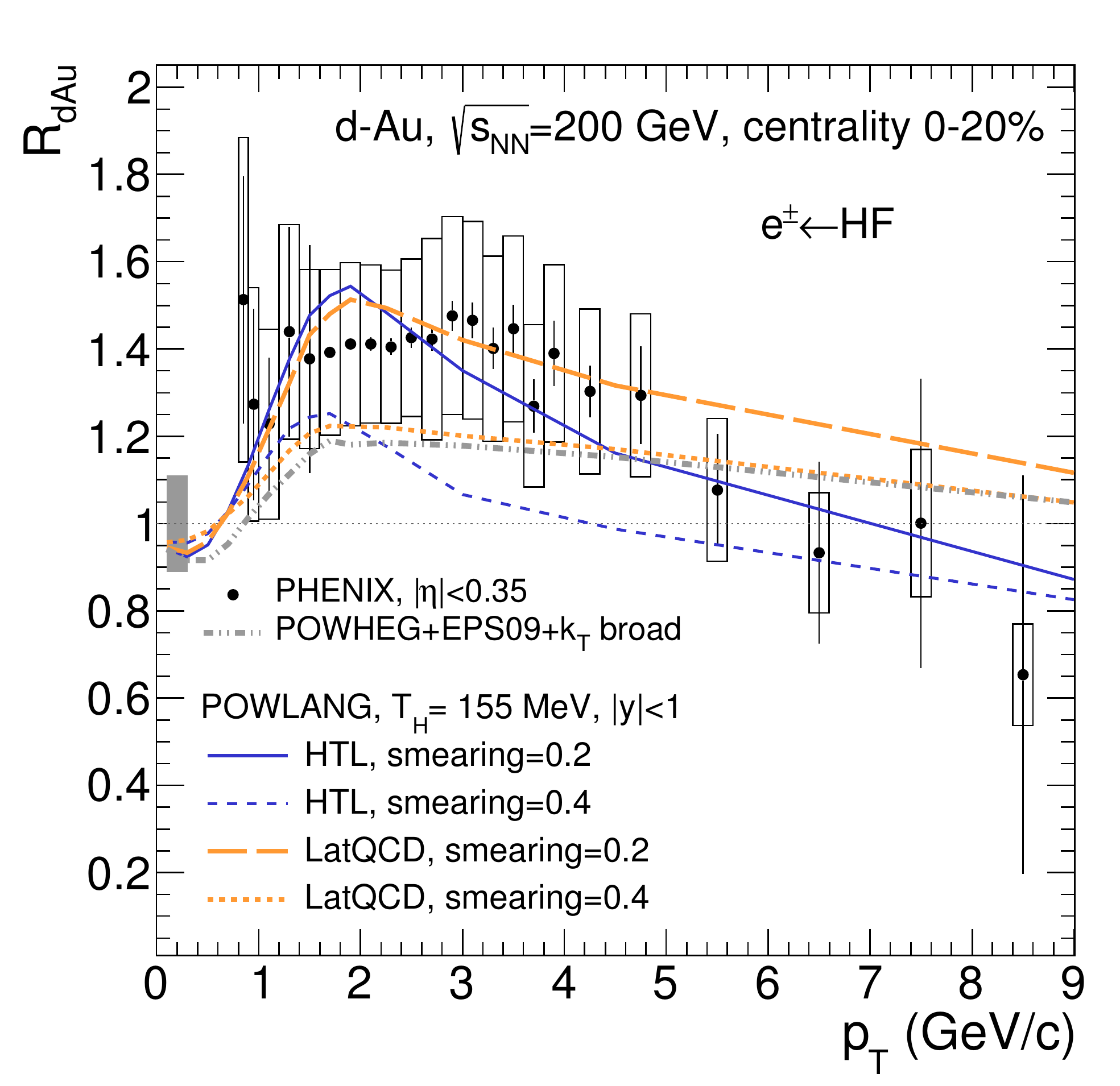}
\includegraphics[width=0.45\textwidth,clip]{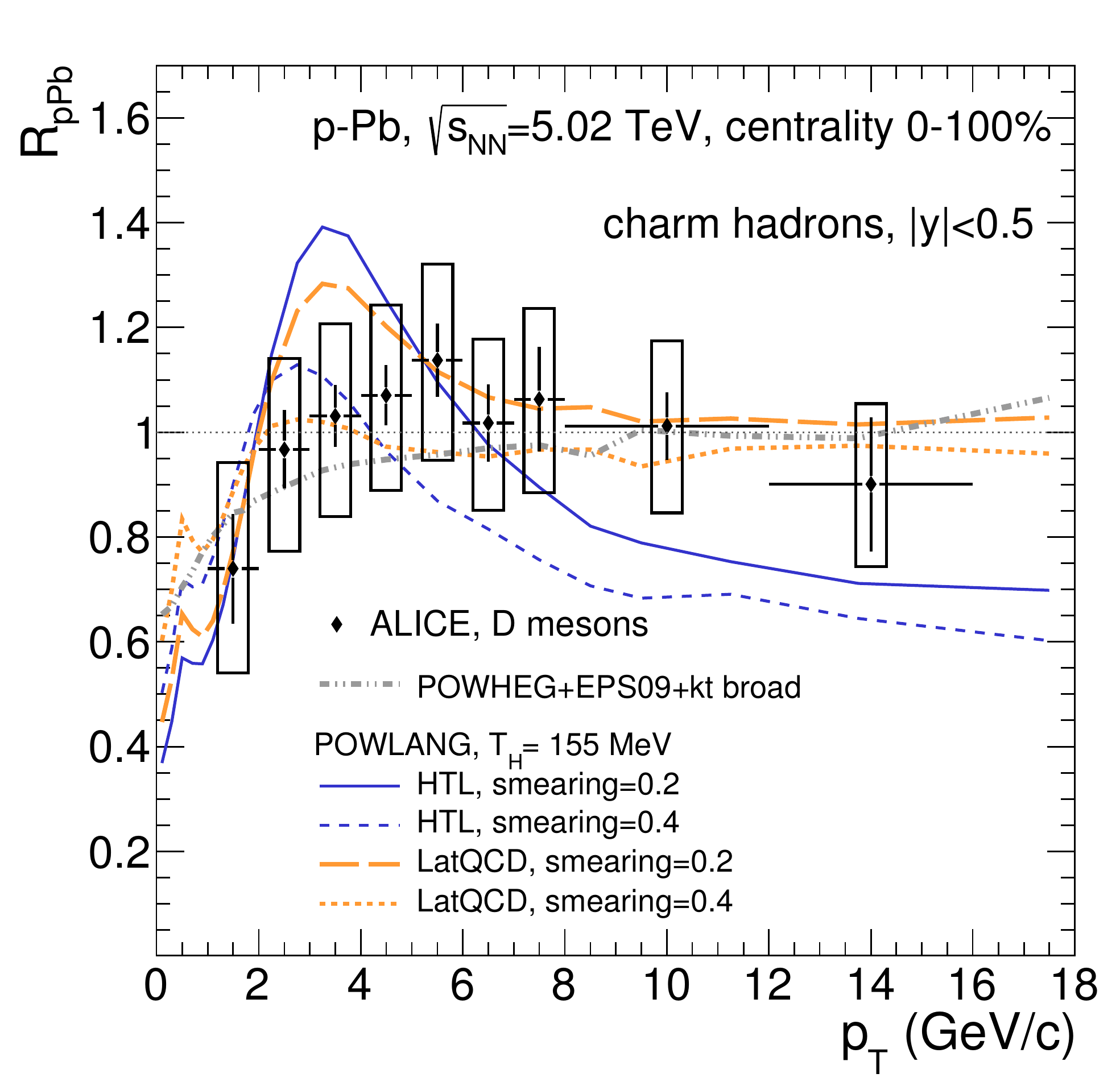}
\caption{Left panel: the nuclear modification factor of HF decay electrons in 0-20\% most central d-Au collisions at RHIC measured by PHENIX~\cite{PHENIX-dAu}. Right panel: the nuclear modification factor of $D$-mesons in 0-100\% p-Pb collisions at the LHC measured by the ALICE collaboration~\cite{ALICE-pPb}. The various curve refers to different choices of transport coefficients and initial conditions.}
\label{small}       
\end{figure}
One of the most surprising findings in the experimental search for the Quark-Gluon Plasma is certainly the signature of possible collective effects, suggestive of the formation of a hot strongly-interacting medium, recently observed in collisions involving small projectiles, like p-Pb at the LHC and d-Au (and now also $^3$He-Au) at RHIC, in particular when selecting events characterized by a high multiplicity of produced particles. Various observables support the above picture, in particular transverse-momentum spectra and azimuthal correlations, suggesting that soft hadrons take part in a common collective expansion, with a radial, elliptic and triangular flow arising from the response of a strongly-interacting medium to the initial pressure gradients and fluctuating geometry. On the contrary, high-$p_T$ hadrons and jets look essentially unmodified with respect to the p-p benchmark. Due to the peculiar role of HF particles, produced in initial hard events but tending to (at least partially) thermalize with the medium in heavy-ion collisions, it is clearly of interest to study their behaviour also in small systems, looking for possible modifications in the final spectra introduced by a hot deconfined plasma. Due to the large systematic error bands, current experimental data are not able to provide a clear answer: the nuclear modification factor of HF electrons (from charm and beauty decays) in d-Au collisions at RHIC measured by PHENIX~\cite{PHENIX-dAu} seems to display a moderate enhancement in the intermediate-$p_T$ region; on the other hand, within the present uncertainties, the $D$-meson $R_{\rm pA}$ measured by ALICE in p-Pb collisions at the LHC looks compatible with unity.
In~\cite{small} the HF transport setup developed in~\cite{epj3} was applied to such small systems. At variance with the nucleus-nucleus case, the initial energy-density profile is not related to the impact parameter of the collision, but to event-by-event fluctuations. Hence, one starts the hydrodynamic evolution of the medium with Glauber Monte-Carlo initial conditions, so to capture at least the fluctuations arising from the random event-by-event nucleon positions. As discussed in detail in~\cite{small}, the final results arise from the interplay of several initial/final-state effects: nuclear PDF's, with gluon shadowing tending to reduce charm production at low-$p_T$; transverse-momentum broadening in nuclear matter (before the initial $Q\overline{Q}$ production), tending to move particles to moderate $p_T$; interaction with the deconfined plasma produced in the collision, tending to suppress heavy-quark spectra at high-$p_T$ and making them inherit part of the flow of the fireball; in-medium hadronization, which also transfer part of the (radial and elliptic) flow of the fireball to the final HF hadrons. Overall, the final outcomes are the ones shown in Fig.~\ref{small}, in which the various curves reflect the different initial conditions and transport coefficients, giving an assessment of the theoretical uncertainty.

\section{Conclusions and perspectives}\label{sec:conlusions}
The comparison between transport calculations and experimental data provides robust indications that heavy quarks strongly interact with the hot deconfined medium formed in high-energy nuclear collisions. Due to such interactions, besides soft hadrons, also heavy-flavour particles carry signatures of the collective flow of the background medium, although the expansion and limited lifetime of the latter may prevent them to reach full kinetic equilibrium with the plasma. 
A number of theoretical questions and experimental challenges remain to be addressed. Charm measurements down to $p_T\to 0$ will be important to get information on the flow and thermalization of charm and its total cross-section in heavy-ion collisions. $D_s$ and $\Lambda_c$ measurements will give a definite answer on possible changes of HF hadronization due to the recombination of the heavy quarks with light thermal partons, which seems currently supported by D-meson $R_{\rm AA}$ and $v_2$ data. Furthermore, $D_s$ and $\Lambda_c$ reconstruction will contribute to a quantitative estimate of the total charm cross-section, relevant for the study charmonium suppression in heavy-ion collisions on solid ground.
In the near future the issue of HF production in small systems (e.g. high-multiplicity p-A collisions), already started in the last few years, will certainly deserve further investigation in order to point out and give a quantitative assessment of initial/final-state effects arising e.g. from the nuclear PDF's and from the possible formation of a hot deconfined medium. Besides the interest for the heavy-ion community, theoretical results for charm production in high-energy p-A collisions validated against experimental data will be of relevance for experiments like IceCube, to get a robust estimate of the charm-decay contribution to the background of high-energy atmospheric neutrinos.

The most important challenge, which would represent a major experimental achievement, remains however the extraction of the HF transport coefficients in hot-QCD: are current experimental data sufficient to put tight constraints on them? In case so, how do they compare with first-principle theoretical calculations? Notice that the concept itself of transport coefficient relies on the idea that the dynamics of the heavy quarks in the medium can be given an effective Langevin description, requiring simply the knowledge of the average squared-momentum exchange per unit time with the plasma. Although current experimental results, once compared to theory calculations, already gives some qualitative guidance, a solid answer to this issue requires waiting for beauty measurements at low $p_T$ via exclusive $B$-meson hadronic decays. Beauty is in fact a better probe of the medium due to its large mass: the latter makes the initial hard production (described by pQCD) under better control, ensures the effective equivalence between the Langevin and the Boltzmann equation and, furthermore, makes the extraction of HF transport coefficients from lattice-QCD simulations conceived for static quarks more meaningful.

\end{document}